\documentclass{article}
\usepackage{amsmath,amssymb,amsfonts}
\usepackage{graphics,graphicx}
\usepackage[unicode,colorlinks,linkcolor=blue]{hyperref}

\title{Inhomogeneous extra space as a tool \\ for the top-down approach} 
\author{Sergey G. Rubin \\
	National Research Nuclear University "MEPhI", \\ (Moscow Engineering Physics Institute),  \\
	N.I. Lobachevsky Institute of Mathematics and Mechanics,\\
	Kazan  Federal  University, \\
	Kremlevskaya  street  18,  420008  Kazan,  Russia\\
	sergeirubin@list.ru}

\begin{document}

	\maketitle

	\begin{abstract}
		In this paper, the top-down approach for the 6-dimensional space has been elaborated.  The connection between the cosmological constant and the extra space metric has been obtained. The metric can be found with the necessary accuracy.
		It is shown that descent from high energies to the low ones leads to the quantum corrections which influence weakly the metric of extra space.
	\end{abstract}

	\section{Introduction}
	Nowadays it becomes more or less clear that the physical laws are formed at high energies where we may only guess about the Lagrangian structure \cite{Brandenberger:2006vv,Tegmark:2005dy}. It is assumed that the values of observable parameters are the result of the evolution of our Universe started at high energies. Observed low-energy physics depends on parameters and initial conditions which have been formed at high energies  \cite{Loeb:2006en,Ashoorioon:2013eia}.
	
	The natural values of the physical parameters are assumed to be quantities of the order of the Planck scale. At the same time, the observed parameter is determined at low energies, and their values are concentrated around electroweak scales and below. The ratio of these two scales is a small parameter, which creates difficulties in constructing the primary theory at high energy. Many attempts have been made to reconcile these two contradictory positions, but the skepticism of the scientific community remains. Quantum corrections only aggravate the situation.
	
In this article two ideas are attracted to  soften the problem. Firstly, the connection between the primary physical parameters and the observable ones is achieved by suitable choice of an extra space metric. Secondly, we determine  parameters of Lagrangian at a high energy scale in the spirit of the Effective Field Theory. In this case, quantum corrections being applied to primary parameters do not spoil the result.
	
One of the aims of the fundamental physics is to postulate a Lagrangian depending on primary parameters and find them using their connection with observational values. Suppose that one managed to obtain a set of relationships
	\begin{equation}\label{glow}
	g_k =\Phi_k (\lambda_i(M)) ,\quad k=1,2,...,K;\quad i=1,2,...,I
	\end{equation}
	between primary parameters  $\lambda_i(M)$ at an energy scale $M$ and the observational parameters $g_k$ (the particle masses, coupling constants and so on) at low energies.
	Solving these equations with appropriate precision, one could determine primary parameters $\lambda_i(M)$ at a chosen scale $M$. The implementation of this plan in its entirety is a matter for the future. Nevertheless, an activity in this direction is observed. In the paper  \cite{Krause} warped geometry is used for the solution of the small cosmological constant problem. The hybrid inflation \cite{LindeHyb} has been developed to avoid the smallness of the inflaton mass. The electron to proton mass ratio is discussed in \cite{Trinhammer}. The seesaw mechanism is usually applied to explain the smallness of neutrino to electron mass ratio \cite{Ibarra}.

	The aim of this paper is to establish and analyze only one connection 
	\begin{equation}\label{LambdaObs}
	\Lambda_{obs}=\Phi(\lambda_i(M))
	\end{equation}
	that ought to be considered as small but necessary part of a 
	future theory. It has been proved  earlier \cite{Rubin:2015pqa, BBDGR} that there exists a set of primary parameters that are responsible for the observable value $\Lambda_{obs}$ of the Cosmological Constant (CC). Here much attention is paid to the problem of quantum corrections.
	
	
	It is shown that the idea on an extra space existence facilitates connection of high energy Lagrangian structure and the low energy one.
	The observational smallness of the CC is used to find the extra space metric.
	
	As a mathematical tool, we  use the effective field theory technique - well-known method for theoretical investigation of the energy dependence of physical parameters \cite{Peskin:1995ev}. In this approach, parameters $\lambda_i (M)$  of the Wilson action are fixed at a high energy scale $M$ and the renormalization flow is used to descend to low energies (the top-down approach) \cite{Burgess:2013ara,Hertzberg:2015bta,Babic:2001vv,Dudas:2005gi}. As is usually stated, the parameters $\lambda_i (M)$ of the Wilson action already contain quantum corrections caused by field fluctuations with energies between the chosen scale $M$ and maximal energy scale, the D-dimensional Planck mass $m_D$ in our case. Therefore the natural value of these parameters is $m_D$ that are usually many orders of magnitude greater than the electroweak scale $v\simeq 100$ GeV. 
	
	The research is based on the multidimensional $f(R)$ gravity. The interest in $f(R)$ theories is motivated by inflationary scenarios starting with the work of Starobinsky \cite{Starobinsky:1980te}. The guiding principle underlying general relativity is the local invariance under coordinate transformations. We may use any invariant combination of quantities invariant under the general coordinate transformations keeping in mind two issues. Firstly, a theory must restore the Einstein-Hilbert action at low energies. Secondly, any gravitational action including the  Einstein-Hilbert one is non-renormalizable and should be considered as an effective theory.
	
	The simplest extension of the gravitation theory is the one containing a function of the Ricci scalar $f(R)$. In the framework of such extension, many interesting results have been obtained. Some viable $f(R)$ models in 4-dim space that satisfies the observable constraints are proposed in Refs. \cite{DeFelice:2010aj,2014JCAP...01..008B,2007CQGra..24.3713S}. 
	Stabilization of extra space as the pure gravitational effect has been studied in \cite{Zhuk,2006PhRvD..73l4019B}. It has been shown recently \cite{Rubin:2015pqa} that the $f(R)$ model with the deformed nonuniform extra space is able to reproduce the 4-dim Minkowski metric. 

	The extra dimensions have now become a widespread tool to obtain new theoretical results \cite{Abbott:1984ba,Chaichian:2000az,Randall:1999vf,Brown:2013fba}.
	The idea of inhomogeneous extra space has been developed in \cite{Gani:2014lka,Rubin:2015pqa,Rubin:2014ffa} and plays one of the central roles in this research.  It influences low energy physics together with physical parameters of a Lagrangian. At the same time
	an accidental formation of manifolds with various metrics and topologies may be considered as a source of different universes whose variety is connected with a continuous set of extra space metrics.
	Entropic mechanism of a metric stabilization is considered in \cite{Kirillov:2012gy}. Stationary extra space metric is the final result of a metric evolution governed by the classical equation of motion, and hence the final stationary metric depends on initial configuration. One could keep in mind an analogy with the black hole mass where the Schwarzschild metric depends on an initial matter distribution.
	In the framework of the scalar-tensor theory Weinberg \cite{Weinberg}  has proved that the firm fine-tuning of initial parameters of a Lagrangian is necessary if metric and scalar fields are constant in space-time. The latter means that the solution of the problem should be seeking in the class of non-uniform configurations of metrics and fields. Metrics of the deformed extra space discussed in this paper belong to this class. 
	
	The plan of the paper consists of three steps. In Section \ref{QCorrMink} we consider a scalar field as the source of quantum corrections to the Lambda term.  It will be shown that they are small relative to primary parameter value at high energies where physical parameters are fixed initially. The appropriate metric of inhomogeneous extra space id discussed in Section \ref{Cceq0}. In Section \ref{QC6D} the scalar field quantum corrections on the inhomogeneous background are analyzed.
	
	\section{Quantum corrections caused by the scalar field. Minkowski space.}\label{QCorrMink}
	
	 The general goal of the top-down approach is to fix primary parameters by comparison with the experimental data at low energies.  According to the Effective Field Theory, quantum fluctuations with energies in the interval $(M, M_{Pl}), M \ll M_{Pl}=1$ had been involved in the parameters of action. It means that their natural values are of the order of the Planck scale. 	Primary parameters are assumed to be formed at high energies, $M$ in our case.

Descending to the electroweak scale $v$ or lower where all observations are used to perform is the necessary step. This process is accompanied by an alternation of physical parameters due to quantum fluctuations. The aim of this section is to demonstrate that quantum corrections are small. The toy model for the scalar field is considered to study the quantum corrections which are the result of integration out the quick modes in the energy interval $ (v, M), v\ll M)$. The scalar field action is written in the standard form
	\begin{eqnarray}\label{act0}
	&&S_{\chi}=\int d^n z \left[\frac12 \partial_{A} \chi g_{n}^{AB}\partial_{B} \chi - U(\chi;\lambda)  -c\right], \\
	&&U(\chi;\lambda)=\lambda_2\chi^2  + \lambda_4\chi^4,\quad 0<\lambda_4(M) \ll \lambda_2(M) \sim c(M)\sim 1 \label{U}
	\end{eqnarray}
	acting in the ordinary 4-dimensional Minkowski space. In this section we study quantum corrections to the parameter $c(M)$ which may be considered as the primary cosmological constant at the scale $M$.

The generating functional for action \eqref{act0} at the scale $M$	\begin{eqnarray}\label{ZM}
	&& Z_0^M=\int_0^M [D\chi]_{M} \exp\left(iS_{\chi}\right).
	\end{eqnarray}
plays the central role in the effective field theory approach.
Here and in the following a subscript and superscript indicate an interval of momentum in the Euclidean space $k_E$ that are taken into account. Thus, functional \eqref{ZM} is the result of integrating out quick modes $M<k_E<m_D$. The D-dimensional Planck mass $m_D$ is considered as the maximal energy scale in the rest of the article. 
	
	Let us integrate out modes with Euclidean momentum $k_E$ in the interval $v<k_E<M$ in generating functional \eqref{ZM} and shift down to the electroweak scale $v\sim 100$ GeV. To this end one should decompose the scalar field as follows
	\begin{eqnarray}\label{quickslow}
	&&\chi(x)=\chi_q(x)+\chi_s(x).
	\end{eqnarray}
	Here quick $\chi_q(x)$ and slow $\chi_s(x)$ modes in 4-dim Euclidean space are \begin{equation}
	\chi_q(x)=\int_{k_E=v}^{k_E=M}\frac{d^4 k_E}{(2\pi)^4}e^{-ik_E x}\chi_{k_E}(x);\quad \chi_s(x)= \int_{k_E=0}^{k_E=v}\frac{d^4 k_E}{(2\pi)^4}e^{-ik_E x}\chi_{k_E}(x)
	\end{equation}
	correspondingly.
	
	Substitution \eqref{quickslow} into \eqref{ZM} gives the generating functional in the form
	\begin{eqnarray}\label{Z0M}
	&& Z_0^M  =Z_0^v\cdot\int_{v}^{M}D\chi_q \exp{\{i\int d^4x\sqrt{g_4}\left[\frac12 (\partial\chi_q)^2 -\lambda_{2}\chi_q^2 -\delta U(\chi_q,\chi_s)\right]\} },\nonumber \\
	&&Z_0^v=\int_{0}^{v}D\chi_s \exp{\{i\int d^4x\sqrt{g_4}\left[\frac{1}{2}(\partial\chi_s)^2-U(\chi_s;\lambda)
		\right]\}}
	\end{eqnarray}
	where 
	\begin{equation}\label{dU}
	\delta U(\chi_q,\chi_s)=4\lambda_{4}\chi_q^3\chi_s  
	+6\lambda_{4}\chi_q^2\chi_s^2
	+ 4\lambda_{4}\chi_q\chi_s^3+ \lambda_{4}\chi_q^4.
	\end{equation} 
	Here we have taken into account orthogonality of $\chi_{s}$ and  $\chi_q$.
	
	{The way to integrate out the field $\chi_q$ from \eqref{Z0M}, provided that the coupling constant $\lambda_{4}$ is small, is well known (see for example textbook \cite{Peskin:1995ev})}. 
	Consider generating functional
	\begin{equation}\label{ZvM}
	Z_v^M=\int_{v}^{M}D\chi_q \exp{\{\int  d^4x\sqrt{g_4}\left[\frac12 (\partial\chi_q)^2 -\lambda_{2}\chi_q^2 -\delta U(\chi_q,\chi_s)+\chi_q(x)J(x)\right]\} }
	\end{equation}
	as a functional of an external current $J$.
	Then the result of integrating out quick modes is as follows
	\begin{eqnarray}
	&&Z_v^M = e^{\epsilon -i\int d^4x\left[ \delta U\left(\frac{-i\delta}{\delta J(x)},\chi_s\right)\right]}
	\cdot e^{\frac{-i}{2} \int d^4x d^4x' J(x)\Delta (x-x')J(x')}, \label{JDJ}\\
	&&\epsilon \equiv -\frac12 Sp\ln \left(\square _4 +  2\lambda_{2} \right)_v^M ,\\
	&& \Delta(x)=\int \frac{d^4k}{(2\pi)^4}\frac{e^{-ikx}}{k^2-2\lambda_{2}+i\epsilon}.
	\end{eqnarray}
	After the Wick rotation quantum correction $\epsilon$ acquires the form
	\begin{equation}\label{delS}
	\epsilon =-i\frac{VT}{16\pi^2} \int_v^M dk_E k_E^3  \ln(k_E^2  + 2\lambda_{2} )
	\end{equation}
	and can be easily calculated. As the result, the contribution \eqref{delS} to the bare cosmological constant $c(M)$
	\begin{equation}\label{du2}
	\delta c =\frac{\epsilon}{iVT} = -\frac{M^4}{64\pi^2} \ln(2\lambda_{2})+o\left({M^4}\right)
	\end{equation}
	is small because of the inequality $M\ll 1$.
	
	The integral in \eqref{delS} is usually estimated keeping in mind that a cutoff scale is much greater than the Lagrangian parameters, i.e., the inequality $M^2\gg \lambda_{2}$ holds, see recent discussion in \cite{Martin}. First estimation has been presented by Ya. Zeldovich in 1967 \cite{ZeldCC}  where the proton mass was used as the maximum energy scale.  In our case the situation is different. Indeed, the scale $M$ is chosen such that $M\ll1$ while a natural value of the effective parameter $\lambda_{2}\sim 1$, see \eqref{U}. In both cases, the corrections are proportional to the fourth power of the energy scale $M$. This is not surprising, since the chosen scale $M$ is still much larger than the electroweak scale $v$.
	
Estimation \eqref{du2} for the quantum corrections $\delta c$ at the scale $M\sim 10^{15}$ GeV gives
	\begin{equation}\label{du2n}
	\delta c \sim 10^{-19} \sim 10^{57} \text{GeV}^4.
	\end{equation}
	This value  is negligibly small as compared to the primary (bare) value $c \sim 1 =(10^{19})^4 GeV^4$ of the $\Lambda$ term,
	\begin{equation}\label{du2m}
	\delta c/c(M) \ll 1
	\end{equation}
	and is huge as compared to the observational value. The latter is not a great problem if our intention is to find values of the physical parameters at a high energy scale. Indeed, quantum corrections must be compared to primary, physical parameters rather than the observational ones.
	
	%
	%
	%
	
	%
	%
	
	It is interesting to check for future studies that correction to the mass $m=\sqrt{2\lambda_{2}}$ also contains the small parameter $M$ and hence is small. To verify this let us estimate quantum corrections produced by terms proportional to $\chi_q^2$.  The latter can be extracted from \eqref{dU}, \eqref{ZvM}  and has the form
	\begin{equation}\label{dU2}
	\delta U_2 \equiv {6\lambda_{4}}\chi_q(x)^2 \chi_s(x)^2.
	\end{equation}
	Receipt \eqref{JDJ} with $\delta U_2 $ instead of $\delta U$ leads to the quantum correction 
	$$\delta U_s(\chi_s)=\int d^4x 6\lambda_{4}\Delta(0) \chi_s(x)^2$$ 
	to the potential in the first multiplier $Z_0^v$ in expression \eqref{Z0M}. Here
	\begin{equation}
	\quad \Delta(0)=-\int
	\frac{d^4 k_E}{(2\pi)^4}\frac{1}{k_E^2 +2\lambda_{2}}\simeq -\frac{1}{64\pi^2} \frac{M^4}{\lambda_{2}}
	\end{equation} 
	for $M^2 \ll \lambda_{2}$. This means that the quantum correction to $\lambda_{2}$ 
	\begin{equation}\label{key}
	\delta\lambda_{2}=6\lambda_{4}\Delta(0)=-\dfrac{3M^4}{32\pi^2}\frac{ \lambda_{4}}{\lambda_{2}}
	\end{equation}
	is small due to the last inequality in \eqref{U} and the choice of energy scale $M\ll 1$. Hence, the quantum correction $\delta m \sim \delta \lambda_{2}$ to the mass $m$ is also small.
	
We can conclude that the quantum corrections are small in comparison with the primary physical parameter. The mass of the scalar particles remains on the order of the Planck scale, which only means that they can not be created at low energies.
A much more serious defect lies in the fact that it is not possible to neutralize the difference between the primary and observational values of CC. We must complicate the model to solve this problem.
	To this end one may draw on the method developed in articles \cite{Rubin:2015pqa} and \cite{BBDGR}. As has been shown in \cite{Rubin:2015pqa}, the problem can be strongly facilitated on the classical level by the 6-dim scalar-tensor gravity with higher derivatives. Moreover, the way of explanation of the CC smallness in the framework of pure gravity without scalar fields was studied in \cite{BBDGR}. The latter is shortly discussed in next Section and the Appendix for clarity.
	
	\section{Inclusion of inhomogeneous extra space}\label{Cceq0}
	In this Section we shortly consider the connection of the CC value and the form of extra space. The discussion is performed on the classical level while the quantum corrections are considered in the next Section. Following the ideas developed in  \cite{Rubin:2015pqa,BBDGR}, consider $f(R)$ gravity with the action
	\begin{eqnarray}\label{act01}
	&&S_g=\frac{m_D^4}{2}\int d^6 z \sqrt{-g_6}f(R), \\
	&& f(R)=R+aR^2 + c \label{f}
	\end{eqnarray}
	acting in a 6-dim space. The constant $c$ that was written explicitly in \eqref{act0} is involved now into the definition of the function $f(R)$.
	
	The metric is assumed to be the direct product $M_4\times V_2$ of the 4-dim space $M_4$ and 2-dim compact space $V_2$
	\begin{equation}\label{interval}
	ds^2 =g_{6,AB}dz^A dz^B = g_{4,\mu\nu}(x)dx^{\mu}dx^{\mu} + g_{2,ab}(y)dy^a dy^b.
	\end{equation} 
	Here $g_{4,\mu\nu}(x)$ and $g_{2,ab}(y)$ are metrics of the manifolds $M_4$ and $V_2$  respectively.  $x$ and $y$ are the coordinates of the subspaces $M_4$ and $V_2$. We will refer to 4-dim space $M_4$ and $2$-dim compact space $V_2$ as the main space and an extra space respectively. The metric has the signature (+ - - - ...), the
	Greek indexes $\mu, \nu =0,1,2,3$ refer to 4-dimensional coordinates. Latin indexes run
	over $a,b = 4, 5$.
	
	There are three scales of energy - the 6-dim Planck mass $m_D$, the characteristic size $r_c$ of the compact extra space $V_2$ and the low energy scale $v\sim 100$GeV. 
	It is assumed that the scales $m_D , r_c^{-1}$ and $v$ satisfy the inequalities
	\begin{equation}\label{ineq}
	v\ll r_c^{-1} \ll m_D =1.
	\end{equation}
	
	The first inequality in \eqref{ineq} means that a characteristic energy scale of extra space is large (the experimental limit is $r_c^{-1} \geq  10^{4}$GeV) and its geometry is stabilized shortly after the Universe creation \cite{2002PhRvD..66d5029N,2002PhRvD..66b4036C,2006PhRvD..73l4019B,Kirillov:2012gy,2007JHEP...11..096G,Abbott:1984ba}. On the other side, quantum behavior dominates at the $m_D$ scale and if one intends to describe a metric of extra space classically, the second inequality must take place. In the following everything is measured in the $m_D$ units.
	
	%
%
	
	As will be shown later condition 
	\begin{equation}\label{vMrc}
	v\ll M \ll r_c^{-1}
	\end{equation}
	for the energy scale $M$ is an appropriate choice. Indeed, the inequality $v\ll M$ permits us to consider the masses of particles be zero. At the same time, excitations of compact extra space  geometry  are known to form the Kaluza-Klein tower with energies $E>r^{-1}_c$. If we start from the energy scale $M\ll r^{-1}_c$, the excitations are suppressed and extra space metric $g_{2,ab}$ represents a stationary configuration described by $(ab)$ part of the classical equations of motion
	\begin{eqnarray} \label{AB}
	&&R_{AB} f' -\frac{1}{2}f(R)g_{6,AB} 
	+ \nabla_A\nabla_B f_R - g_{6,AB} \square f' =0 , \\
	&& \text{+ additional conditions} \nonumber
	\end{eqnarray}
	where $\Box f_R = \nabla^A \nabla_A f_R$.
	
Let us assume the metric of our 4-dim space be the Minkowski metric, $g_{4} = diag (1,-1,-1,-1)$.  The compact 2-dim manifold is supposed to be parameterized by the two spherical angles $y_1=\theta$ and $y_2=\phi$ $(0 \leq\theta \leq \pi, 0 \leq \phi < 2\pi)$.
	The choice of the extra space metric is as follows
	\begin{equation}\label{metric2}
	g_{2,\theta\theta} = -r(\theta)^2;\quad g_{2,\phi \phi}= -r(\theta)^2 \sin^2(\theta).
	\end{equation}
	There is continuous set of extra space metrics - solutions to the differential equations \eqref{AB} - characterized by additional conditions.  Maximally symmetrical extra spaces that are used in great majority of literature represent a small subset of this set. 
	As the additional conditions let us fix the metric at the point $\theta =\pi$ 
	\begin{equation}\label{bound2}
	r(\pi)=r_{\pi} ; \quad r'(\pi)=0 ; \quad  R(\pi)=R_{\pi} ; \quad R'(\pi)=0.
	\end{equation}
	The system of equations \eqref{AB} together with these conditions completely determine the form of extra space metric.
	
	\begin{figure}[h]
		\begin{minipage}[h]{0.32\linewidth}        
			\center{\includegraphics[width=1.\linewidth]{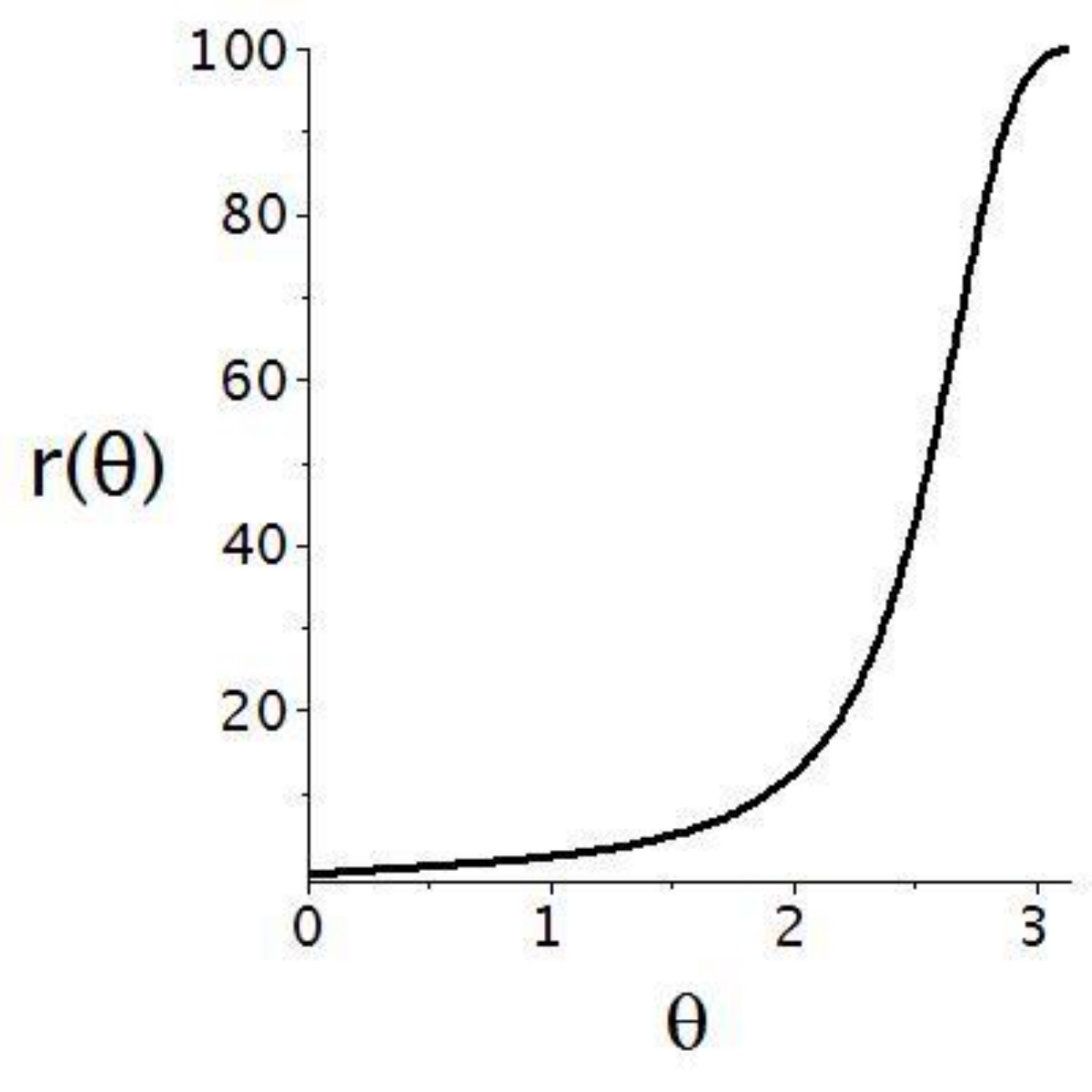}}
		\end{minipage}
		\hfill
		\begin{minipage}[h]{0.34\linewidth}
			\center{\includegraphics[width=1.4\linewidth]{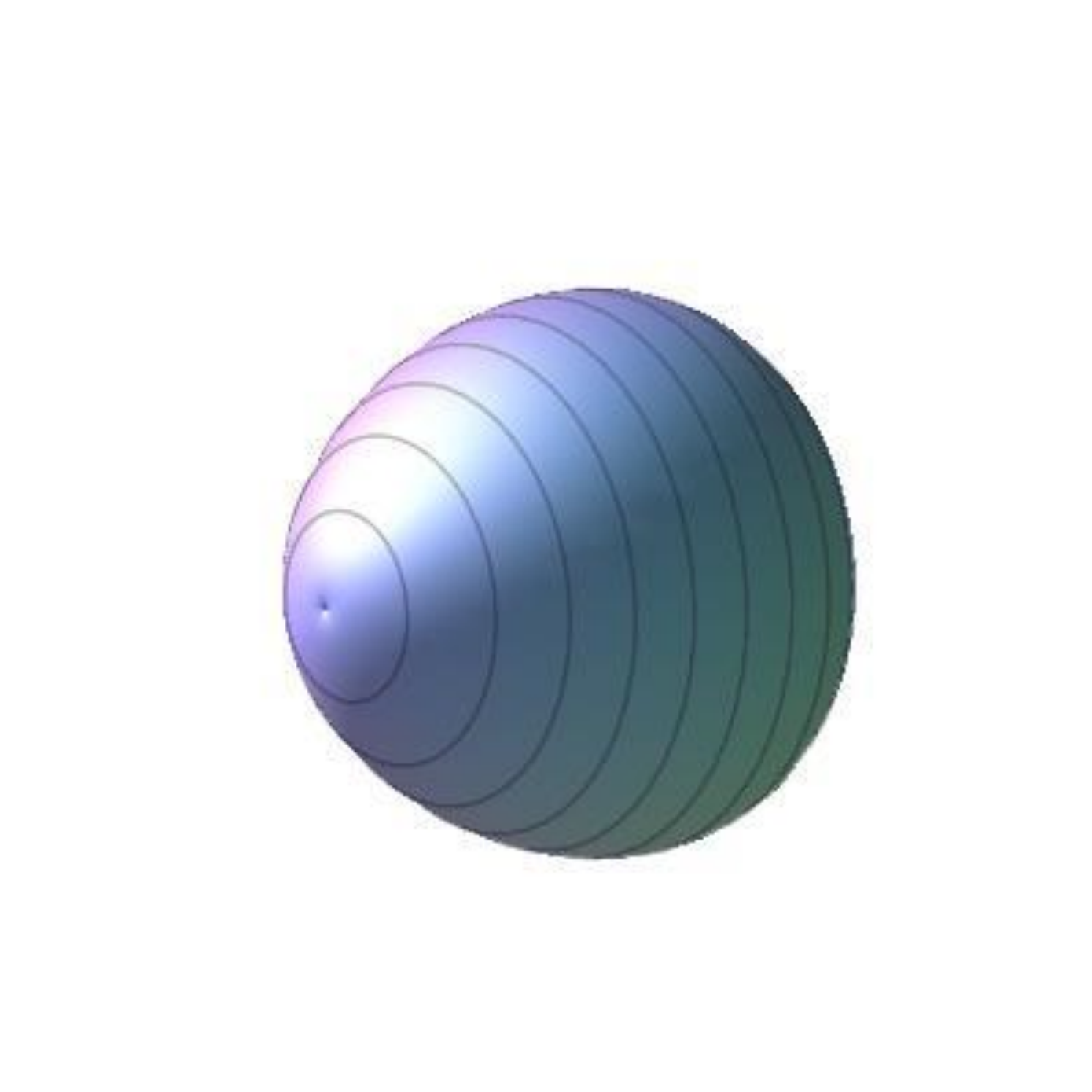}}
		\end{minipage}
		\hfill
		\begin{minipage}[h]{0.32\linewidth}
			\center{\includegraphics[width=0.8\linewidth]{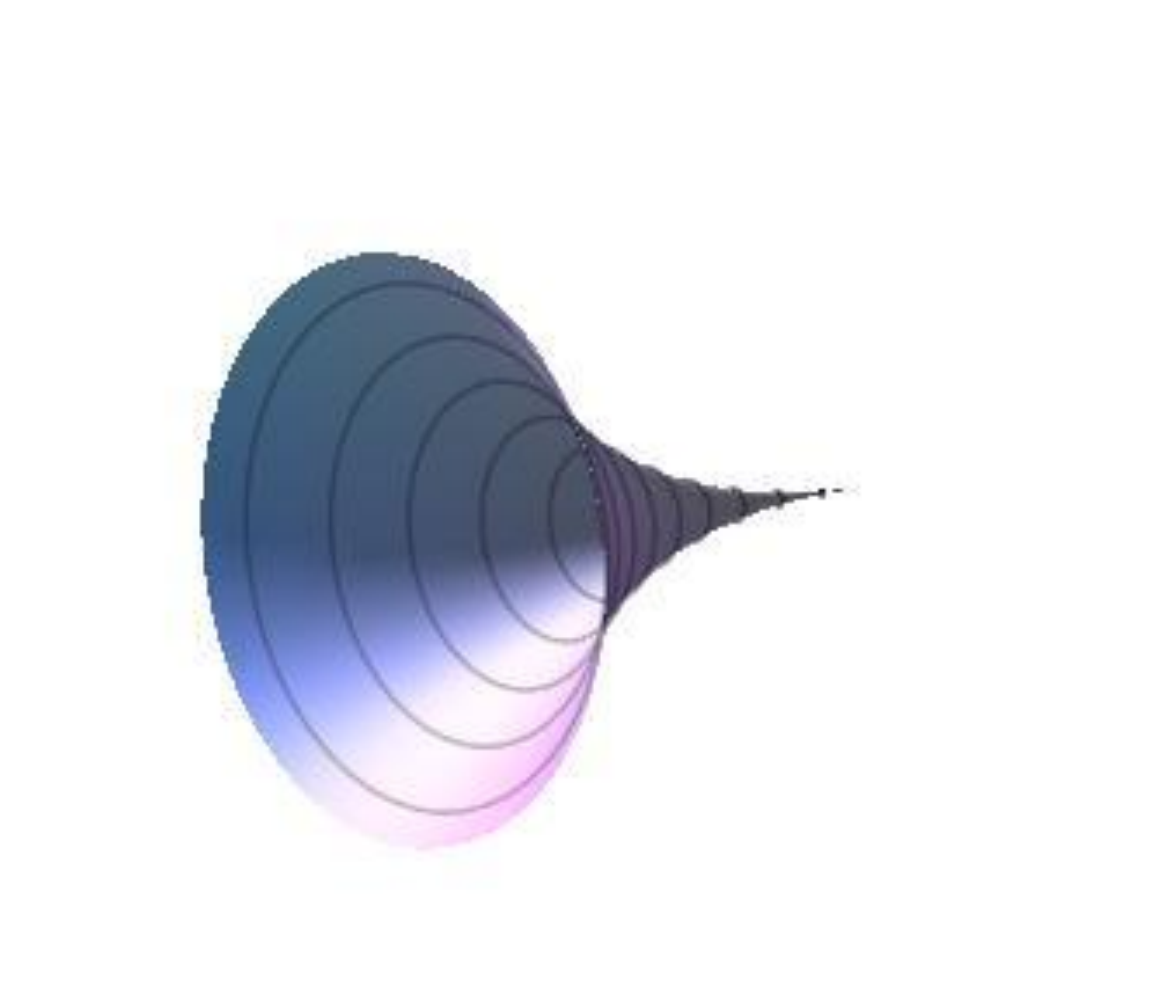}}
		\end{minipage}
		\\
		\hspace*{2cm} (a) \hspace{3.9cm} (b)  \hspace{2.6cm} (c)  
		\caption{
			(a) Radius $r(\theta)$ of extra space for the parameter  
			values $a=-100$, $c = -2.1 \cdot 10^{-3}$ and additional conditions 
			$r(\pi) = 100$, $R(\pi) = 1.0251\cdot 10^{-3}$.  
			(b) 3D plot of the solution. (c) A small part of the  3D plot near   $\theta=0$ (left ``end'').
		}
		\label{form}
	\end{figure}

	Numerical solutions to equation \eqref{AB} with additional conditions \eqref{bound2} are discussed in \cite{Gani:2014lka,Rubin:2015pqa}. It has been found that due to high nonlinearity of the equation, the gravity can trap itself in a small region around $\theta =0$ even without matter contribution. 
	

Next step consists of finding an appropriate 2-dim extra metric with the help of the observable value of CC. General connection is represented in the Appendix, formula \eqref{Lambda}
%
%
%
	It should be stressed that our aim is not to calculate CC with extremal accuracy $10^{-123}$ but to find physical parameters at high energy scale $M$. In this case the left hand side of equation \eqref{Lambda} can be safely substituted by zero and we arrive to the following connection between the physical parameters
	\begin{equation}\label{Lambda0}
	\Lambda_{theor}(a,c,r(\pi),R(\pi)) \equiv - \frac{\pi}{M^2 _{Pl}}\int d\theta \sqrt{|g_2(\theta)|} f(R_2(\theta))\simeq 0.   
	\end{equation}
	
	To be more specific, suppose that the primary parameters of the Lagrangian and the extra space size dictated by the parameter $r(\pi)$ are known
	\begin{eqnarray}\label{param}
	a=-100, c=-2.1\cdot 10^{-3}, r(\pi)=100.
	\end{eqnarray}
	Then numerical solution to equation \eqref{Lambda0} with respect to the Ricci scalar $R(\pi)$ can be obtained,
	\begin{equation}\label{Rpi}
	R(\pi)\simeq 1.0251\cdot 10^{-3}.   
	\end{equation}
 Relative accuracy $10^{-4}$ of this result can be improved if necessary.
	Thus we have found the extra space metric, see  Fig.\ref{form}, of our toy model with appropriate precision.
	
	The 4-dim Planck mass can be found numerically according to expression \eqref{MPl} and equals $M_{Pl}=34 m_D$. This gives the value of the D-dim Planck mass, 
	$
	m_D \simeq 3\cdot 10^{17}GeV.
	$
	Therefore the scale $M$ may be chosen in the interval $10^2 \ll M \ll 10^{17}$GeV, see inequality \eqref{vMrc}.

	The intermediate conclusion is that the smallness of CC can be used at the classical level for fixing appropriate metric of the inhomogeneous extra space. In the next section, we discuss the role of quantum corrections and their influence on this classical result.
	
	\section{Quantum corrections caused by the scalar field. Inhomogeneous extra space background.}\label{QC6D}
	In this section, we discuss the way to integrate out the extra space coordinates in expression \eqref{act0} to obtain an effective 4-dim action describing physics at the scale $M$. Information about the extra space metric will be stored in the effective parameters at this scale.
	
	As discussed below  formula \eqref{vMrc}, the excitations of the extra space metric are suppressed due to the choice of the scale $M$. The same arguments may be applied to the scalar field excitations on the 2-dim extra space. Classical distribution can be obtained from the equation of motion
	\begin{equation}\label{eqn6}
	\square_{6}\varphi+U'(\varphi;\lambda)=0,
	\end{equation}
	where $\square_{6}$ is 6-dim d'Alemert operator.
	Let us decompose the scalar field into a series of orthonormal functions $Y_n$ acting on the extra space. 
	The smallness of  fluctuations at the scale $M$ means that we may limit ourselves to the first term
	\begin{equation}\label{xY}
	\varphi(x,\theta) =\chi (x) Y(\theta),
	\end{equation}
	where $Y(\theta)$ is a solution to classical equation
	\begin{equation}\label{eqY1}
	\square_2 Y(\theta) +2\lambda_2 Y(\theta)=0.
	\end{equation}
	This equation is obtained from \eqref{eqn6} by neglecting small terms containing $x-$ derivatives, see Appendix formula \eqref{lll}, and those terms proportional to the small coupling $\lambda_4$.
	More explicit form of this equation
	\begin{equation}\label{eq3}
	\cot(\theta)\partial_{\theta}Y(\theta) + \partial^2 _{\theta}Y(\theta)  - 2\lambda_2 r (\theta)^2Y(\theta) =0
	\end{equation} 
	can be obtained by substituting  metric \eqref{metric2} into equation \eqref{eqY1}. Approximate solution to this equation
	\begin{equation}\label{WKBsol}
	Y(\theta)=Ce^{-\sqrt{2\lambda_2}\int_0^{\theta}d\theta'r(\theta')}
	\end{equation}
	has been found in \cite{Rubin:2015pqa}. Here 
	$$C=\left[2\pi \int d\theta r(\theta)^2 \sin\theta e^{-2\sqrt{2\lambda_2}\int_0^{\theta}d\theta'r(\theta')}   \right]^{-1/2}$$ 
	is the normalization constant.
	
	After substitution \eqref{xY} into expression \eqref{act0} we get the following form of the effective 4-dim action for the gravity with the scalar field
	\begin{eqnarray}
	&&S=S_g +S_{\chi} \nonumber \label{act1}\\
	&&S_{\chi}=\frac{1}{2}\int d^4  x \left[ \frac{1}{2}\partial_{\mu}\chi(x)g^{\mu\nu}\partial_{\nu}\chi(x) - U(\chi;\lambda_{eff}) \right] , \label{Ssc}
	\end{eqnarray}
	where
	\begin{eqnarray}\label{LLa}
	&&\lambda_{eff,2}=\lambda_2-\frac{1}{2}\int d\theta \sqrt{|g_2(\theta)|} \partial_aY(\theta)g_2^{ab}(\theta)\partial_b  Y(\theta) \simeq 3\lambda_2 , \label{l2}\\
	&&\lambda_{eff,4} = \lambda_4\cdot \int d\theta \sqrt{|g_2(\theta)|} Y(\theta)^4  \label{l4}
	\end{eqnarray}
	and $S_g$ is the gravitational action \eqref{act01}. As was shown in Section \ref{Cceq0}, the extra space metric may be chosen such that the metric of our 4-dim space is arbitrary close to the Minkowskian metric. The second equality in \eqref{l2} is true due to the form of solution \eqref{WKBsol} and metric $g_2^{\theta\theta}(\theta)=-r(\theta)^{-2}$.
	New effective parameters $\lambda_{eff,2,4}$ depend on the functions $Y(\theta), g_{2,ab}(\theta)$ and, hence, on the additional conditions $r(\pi), R (\pi)$. 
	A connection of effective 4-dim parameters with the metric of extra space is the well-known result. The most known example is connection of the Planck mass to a D-dim Planck mass \cite{2002PhRvD..66d5029N}
	$M_{Pl}^2=m_D^{D-2}V_e,$
	where $V_e$ stands for an extra space volume.
	
	The mass of the field $m_{eff}=\sqrt{2\lambda_{eff,2}}$ remains of the same order of the magnitude, $m_{eff}\sim m=\sqrt{2\lambda_{2}}$. Specific example considered in Section 6 gives $m\simeq 3\cdot 10^{17}$ GeV. To perform numerical calculation of the coupling constant $\lambda_4$, let us use the metric presented in Fig.1. In this case the integral in \eqref{l4} can be evaluated and we obtain the renormalized parameter
	\begin{equation}\label{leff4}
	\lambda_{eff,4}\simeq 0.19\lambda_{4} .
	\end{equation}
	Therefore, the effective parameters satisfy the same inequalities as those mentioned in \eqref{U}, $\lambda_{eff,4}\ll \lambda_{eff,2}\sim 1$.
	
	We have restored the scalar field action \eqref{act0} with the effective parameter values \eqref{l2} and \eqref{l4} depending on the extra space metric and the values of primary parameters.   Therefore, the result of Section \ref{QCorrMink} may be applied to the action \eqref{act1} to study the influence of quantum corrections. 
	
	The inclusion of the quantum corrections means that the bare value of the parameter $c$ should be substituted by $c + \delta c$ in equation \eqref{Lambda0}:
	\begin{equation}\label{LambdaQC}
	\Lambda_{theor}(a,c+\delta c,r(\pi),R(\pi)) =0.
	\end{equation}
	
	As was discussed above, see \eqref{du2m}, the ratio $\delta c/c$ is small. Therefore the shift $c\rightarrow c+\delta{c}$ can be easily compensated by a small shift in the Ricci scalar $R(\pi)\rightarrow {R}(\pi) +\delta R(\pi)$ in equation   \eqref{LambdaQC}. This means that the quantum corrections lead only to a  more accurate determination of the extra space metric depending on $R(\pi)$ in our case.
	
\section{Conclusion}
	
	In this paper, the top-down approach for the 6-dim space has been elaborated.  On the basis of toy model, the connection between the observable 4-dim Lambda term and the extra space metric $g_2$ has been obtained. It permits to find this metric with an appropriate precision provided that other parameters are known.
	
Values of the physical parameters $\lambda_i(M)$ where chosen of the order of unity in 6-dim Planck units $m_D$ at the high energy scale. It is shown here that the descent from the high energy scale $M$ to the electroweak scale $v$ leads to the quantum corrections $\delta\lambda_i(M)$ which are small as compared to the primary (bare) values $\lambda_i(M)$. For specific case discussed in the article,  the quantum corrections to the primary $\Lambda$ term permit to calculate metric of the extra space more accurately.
		


	The results of this research are valid in the energy interval  $v\ll M \ll m_D$.  If $M\sim m_D$ the quantum corrections are compatible with the classical results so that their influence cannot be controlled. If $M\lesssim v$ knowledge of the observable particle masses is necessary to evaluate integrals like those presented in \eqref{JDJ} and \eqref{delS}.
	
	The model discussed above contains the set of primary parameters $a,c$, $r(\pi),R(\pi)$. Connections for other low energy physical parameters similar to \eqref{Lambda} should be included if one wishes to determine all the primary parameters.  This is the subject of future research. 
	The discovery of the gravitational waves (GW) \cite{Abbott2016} provides an additional tool for such activity.  
	
	GW propagate from distant galaxies to the Earth with practically no distortion caused by interaction with the matter.  In this regard, GW become a significant tool in the analysis of extra dimension properties. In the near future, some restrictions on the theory of gravity can be put when the number of GW sources will amount to several hundred. Though many interesting results for sure will be obtained one can foresee serious difficulties caused by a large amount of models. The problem is that results depend on a structure of extra space, a number of extra dimensions and their size. For example, interesting result has been obtained in \cite{GWEMW} where the difference between  propagations of GW and electromagnetic waves has been studied. This study is applicable only for a one dimensional extra space.  
	
The situation is aggravated by the fact that a lot of gravity theories other than the Einstein-Hilbert theory have been developed up to now. Specific choice of theory could produce additional effects like GW propagation with a speed different from the speed of light \cite{Bettoni}. 
Nevertheless, new methods will undoubtedly help to extract promising directions in gravitational physics. For example, the methodologies that depend on redshift \cite{Bellido} or the ones that are valid at short distances for $z\ll1$\cite{Calabrese} have been elaborated.  Also, substantial review can be found in \cite{Berti}.
	
	\section{Appendix}
	
	Here the explicit form of relationship \eqref{Lambda0} between the Lambda term and the primary parameters \eqref{f} is found. 	
	Throughout the paper, the metric of our 4-dim space is the (almost) Minkowski one, $g_{4} \cong diag (1,-1,-1,-1)$. It means that the Ricci scalar $R_4$ of our space is small compared to the Ricci scalar $R_2$ of the extra space.  More definitely, the inequalities
	\begin{equation}\label{lll}
	R_4 \ll R_2 , \quad \partial_{\mu} \ll \partial_a ,\quad \mu=0,1,2,3,\quad a=4,5
	\end{equation}
	take place if one takes into account the experimental limit on the extra space size $L_n < 10^{-18}$ cm and connection $R_2 \sim 1/L_2^2$ between the size $L_2$ and the Ricci scalar of the extra space. More discussion may be found in \cite{2006PhRvD..73l4019B,2007CQGra..24.1261B}.
	
Let us transform the gravitational part of the action. 	Using inequality \eqref{lll} the Taylor expansion $f(R)=f(R_4 + R_2)\simeq f(R_2)+f'(R_2)R_4 $ in expression \eqref{act0} gives \cite{2006PhRvD..73l4019B}
	\begin{equation}\label{Sg}
	S_g=\pi \int d^4x d\theta \sqrt{|g_4(x)g_2(\theta)|}\left[ R_4f'(R_2(\theta)) + f(R_2(\theta)) \right]. 
	\end{equation}
	Comparison of expression \eqref{Sg} with the Einstein-Hilbert action
	\begin{equation}\label{SEH}
	S_{EH}=\frac{M^2 _{Pl}}{2} \int d^4x  \sqrt{|g_4(x)|}(R_4-2\Lambda_{obs})
	\end{equation}
	leads to the relationship
	\begin{equation}\label{Lambda}
	\Lambda_{obs}=\Lambda_{theor}(a,c,r(\pi),R(\pi)) \equiv - \frac{\pi}{M^2 _{Pl}}\int d\theta \sqrt{|g_2(\theta)|} f(R_2(\theta))
	\end{equation}
	that connects the observable value $\Lambda_{obs}$ of the CC with four main parameters of the model - $a, c$ from expression \eqref{f} and the parameters $r(\pi), R(\pi)$ characterizing the extra space metric \eqref{bound2}. Here $M_{Pl}$ is the 4-dim Planck mass.
	The relationship \eqref{Lambda} represents the particular case of connection \eqref{Lambda0}. Its r.h.s. depends also on a stationary geometry $g_{2,ab} (\theta)$  and hence on the flexible parameters $R(\pi)$ and $r(\pi)$. Comparison of expression \eqref{Sg} with the Einstein-Hilbert action \eqref{SEH} gives also the Planck mass 
	\begin{equation}\label{MPl}
	M^2 _{Pl}=2\pi \int d\theta \sqrt{|g_2(\theta)|}f' (R_2 (\theta) )
	\end{equation}
	as the function of the primary parameters and geometry of the extra space.

	\section{Acknowledgment}
	The author is grateful to I. Buchbinder, D. Kazakov for interesting discussions and O. Moiseeva for valuable remarks.
	This work was supported by the MEPhI Academic Excellence Project (agreement with the Ministry of Education and Science of the Russian Federation of August 27, 2013, project no. 02.a03.21.0005) and by the Russian Government Program of Competitive Growth of Kazan Federal University. 
	The work was supported by the Ministry of Education and Science of the Russian Federation, Project No.~3.472.2017/K.


\begin{thebibliography}{10}
		
		\bibitem{Brandenberger:2006vv}
		Robert~H. Brandenberger, Ali Nayeri, Subodh~P. Patil, and Cumrun Vafa.
		\newblock {String gas cosmology and structure formation}.
		\newblock {\em Int. J. Mod. Phys.}, A22:3621--3642, 2007.
		
		\bibitem{Tegmark:2005dy}
		Max Tegmark, Anthony Aguirre, Martin Rees, and Frank Wilczek.
		\newblock {Dimensionless constants, cosmology and other dark matters}.
		\newblock {\em Phys. Rev.}, D73:023505, 2006.
		
		\bibitem{Loeb:2006en}
		Abraham Loeb.
		\newblock {An Observational Test for the Anthropic Origin of the Cosmological
			Constant}.
		\newblock {\em JCAP}, 0605:009, 2006.
		
		\bibitem{Ashoorioon:2013eia}
		Amjad Ashoorioon, Konstantinos Dimopoulos, M.~M. Sheikh-Jabbari, and Gary Shiu.
		\newblock {Reconciliation of High Energy Scale Models of Inflation with
			Planck}.
		\newblock {\em JCAP}, 1402:025, 2014.
		
		\bibitem{Krause} 
		Axel Krause, 
		Journal of High Energy Physics \textbf{09},  016 (2003).
		
		\bibitem{LindeHyb}
		Linde, Andrei
		Phys. Rev. D \textbf{49}, 15 1994, pp.748-754
		
		\bibitem{Trinhammer}    
		Trinhammer, Ole L.
		Europhysics Letters, \textbf{102} 4,  42002 (2013).
		
		\bibitem{Ibarra}
		Ibarra, A., Molinaro, E., Petcov, S. T.
		Physical Review D \textbf{84}, 013005 (2011).
		
		
		
		\bibitem{Rubin:2015pqa}
		Sergey~G. Rubin.
		\newblock {Scalar field localization on deformed extra space}.
		\newblock {\em Eur. Phys. J.}, C75(7):333, 2015, e-Print: arXiv:1503.05011
		
		\bibitem{BBDGR} 
		K.A. Bronnikov, R.I. Budaev, A.V. Grobov, A.E. Dmitriev, Sergey G. Rubin, 
		JCAP 1710 (2017) no.10, 001, e-Print: arXiv:1707.00302
		
\bibitem{Abbott2016}		B.P. Abbott, et al., Phys. Rev. Lett. 116(6), 061102 (2016). arXiv:1602.03837

		\bibitem{Peskin:1995ev}
Michael~E. Peskin and Daniel~V. Schroeder.
\newblock {\em {An Introduction to quantum field theory}}.
\newblock 1995.
		
	\bibitem{Martin} J. Martin, 
	Everything You Always Wanted To Know About The Cosmological
	Constant Problem (But Were Afraid To Ask), Comptes Rendus Physique 13 (2012) 566-665, arXiv:1205.3365v1  
		
\bibitem{ZeldCC}	Ya. B. Zeldovich,  Cosmological constant and elementary particles,  JETP Lett. 6, 316 (1967)
		
		
		\bibitem{Sahni:1999gb}
		Varun Sahni and Alexei~A. Starobinsky.
		\newblock {The Case for a positive cosmological Lambda term}.
		\newblock {\em Int. J. Mod. Phys.}, D9:373--444, 2000.
		
		\bibitem{Burgess:2013ara}
		C.~P. Burgess.
		\newblock {The Cosmological Constant Problem: Why it's hard to get Dark Energy from Micro-physics}.
		\newblock In {\em {Proceedings, 100th Les Houches Summer School: Post-Planck
				Cosmology: Les Houches, France, July 8 - August 2, 2013}}, pages 149--197,
		2015.
		
		\bibitem{Hertzberg:2015bta}
		Mark~P. Hertzberg and Ali Masoumi.
		\newblock {Can Compactifications Solve the Cosmological Constant Problem?}
		\newblock {\em JCAP}, 1606(06):053, 2016.
		
		\bibitem{Babic:2001vv}
		A.~Babic, B.~Guberina, R.~Horvat, and H.~Stefancic.
		\newblock {Renormalization group running of the cosmological constant and its
			implication for the Higgs boson mass in the standard model}.
		\newblock {\em Phys. Rev.}, D65:085002, 2002.
		
		\bibitem{Dudas:2005gi}
		E.~Dudas, C.~Papineau, and V.~A. Rubakov.
		\newblock {Flowing to four dimensions}.
		\newblock {\em JHEP}, 03:085, 2006.
		
		\bibitem{Starobinsky:1980te}
		Alexei~A. Starobinsky.
		\newblock {A New Type of Isotropic Cosmological Models Without Singularity}.
		\newblock {\em Phys. Lett.}, B91:99--102, 1980.
		
		
		\bibitem{DeFelice:2010aj}
		Antonio De~Felice and Shinji Tsujikawa.
		\newblock {f(R) theories}.
		\newblock {\em Living Rev. Rel.}, 13:3, 2010.
		
		\bibitem{2014JCAP...01..008B}
		K.~{Bamba}, A.~N. {Makarenko}, A.~N. {Myagky}, S.~{Nojiri}, and S.~D.
		{Odintsov}.
		\newblock {Bounce cosmology from F(R) gravity and F(R) bigravity}.
		\newblock {\em J. Cosmol. Astropart. Phys.}, 1:8, January 2014.
		
		\bibitem{2007CQGra..24.3713S}
		L.~M. {Soko{\l}owski}.
		\newblock {Metric gravity theories and cosmology: II. Stability of a ground    state in f(R) theories}.
		\newblock {\em Class and Quantum Grav.}, 24:3713--3734, 2007.
		
		\bibitem{Zhuk}
		{{G{\"u}nther}, U. and {Zhuk}, A. and {Bezerra}, V.~B. and {Romero}, C.},
		{AdS and stabilized extra dimensions in multi-dimensional gravitational models with nonlinear scalar curvature terms R$^{-1}$ and R$^{4}$}, Class. Quant. Grav., 22:3135, 2005, {arXiv:hep-th/0409112},
		
		
		\bibitem{2006PhRvD..73l4019B}
		K.~A. {Bronnikov} and S.~G. {Rubin}.
		\newblock {Self-stabilization of extra dimensions}.
		\newblock {\em Phys. Rev.~\bf{D}}, 73(12):124019, June 2006.
		
		\bibitem{Weinberg}
		Weinberg S. The cosmological constant problem // Rev.Mod.Phys. 1989. 61. 1.
		
		\bibitem{Abbott:1984ba}
		Richard~B. Abbott, Stephen~M. Barr, and Stephen~D. Ellis.
		\newblock {Kaluza-Klein Cosmologies and Inflation}.
		\newblock {\em Phys. Rev.}, D30:720, 1984.
		
		\bibitem{Chaichian:2000az}
		Masud Chaichian and Archil~B. Kobakhidze.
		\newblock {Mass hierarchy and localization of gravity in extra time}.
		\newblock {\em Phys. Lett.}, B488:117--122, 2000.
		
		\bibitem{Randall:1999vf}
		Lisa Randall and Raman Sundrum.
		\newblock {An Alternative to compactification}.
		\newblock {\em Phys. Rev. Lett.}, 83:4690--4693, 1999.
		
		\bibitem{Brown:2013fba}
		Adam~R. Brown, Alex Dahlen, and Ali Masoumi.
		\newblock {Compactifying de Sitter space naturally selects a small cosmological constant}.
		\newblock {\em Phys. Rev.}, D90(12):124048, 2014.
		
		\bibitem{Gani:2014lka}
		Vakhid~A. Gani, Alexander~E. Dmitriev, and Sergey~G. Rubin.
		\newblock {Deformed compact extra space as dark matter candidate}.
		\newblock {\em Int. J. Mod. Phys.}, D24:1545001, 2015.
		
		
		
		\bibitem{Rubin:2014ffa}
		S.~G. Rubin.
		\newblock {The role of initial conditions in the universe formation}.
		\newblock {\em Grav. Cosmol.}, 21:143--151, 2015.
		
		\bibitem{Kirillov:2012gy}
		A.~A. Kirillov, A.~A. Korotkevich, and S.~G. Rubin.
		\newblock {Emergence of symmetries}.
		\newblock {\em Phys. Lett.}, B718:237--240, 2012.
		
		\bibitem{2002PhRvD..66d5029N}
		S.~{Nasri}, P.~J. {Silva}, G.~D. {Starkman}, and M.~{Trodden}.
		\newblock {Radion stabilization in compact hyperbolic extra dimensions}.
		\newblock {\em Phys. Rev.~\bf{D}}, 66(4):045029, August 2002.
		
		\bibitem{2002PhRvD..66b4036C}
		S.~M. {Carroll} et~al.
		\newblock {Classical stabilization of homogeneous extra dimensions}.
		\newblock {\em Phys. Rev.~\bf{D}}, 66(2):024036, July 2002.
		
		
		\bibitem{2007JHEP...11..096G}
		B.~{Greene} and J.~{Levin}.
		\newblock {Dark energy and stabilization of extra dimensions}.
		\newblock {\em Journal of High Energy Physics}, 11:96, November 2007.
		
		\bibitem{2007CQGra..24.1261B}
		K.~A. {Bronnikov}, R.~V. {Konoplich}, and S.~G. {Rubin}.
		\newblock {The diversity of universes created by pure gravity}.
		\newblock {\em Classical Quant. Grav.}, 24:1261--1277, March 2007.    
		
				
		\bibitem{Abbott2016}		B.P. Abbott, et al., Phys. Rev. Lett. 116(6), 061102 (2016). arXiv:1602.03837
		
\bibitem{GWEMW}
Hao Yu, Bao-Min Gu, Fa Peng Huang, Yong-Qiang Wang,  Xin-He Meng, Yu-Xiao Liu,
Probing extra dimension through gravitational wave observations of compact binaries and their electromagnetic counterparts, JCAP 1702 (2017) no.02, 039.

\bibitem{Bettoni}
Dario Bettoni, Jose Mar?a Ezquiaga, Kurt Hinterbichler, Miguel Zumalac?rregui, Speed of Gravitational Waves and the Fate of Scalar-Tensor Gravity, Phys.Rev. D95 (2017) no.8, 084029
	
\bibitem{Bellido}	Gravitational wave source counts at high redshift and in models with extra dimensions
	Juan Garcia-Bellido, Savvas Nesseris, Manuel Trashorras, JCAP 1607 (2016) no.07, 021
		
		
\bibitem{Calabrese}	E. Calabrese, N. Battaglia, and D. N. Spergel, “Testing Gravity with Gravitational Wave Source Counts”, Class.Quant.Grav. 33 (2016) no.16, 165004, arXiv:1602.03883		

		
\bibitem{Berti}	Emanuele Berti, Kent Yagi, Nicolas Yunes,	Extreme Gravity Tests with Gravitational Waves from Compact Binary Coalescences: (I) Inspiral-Merger,
		e-Print: arXiv:1801.03208 
		
		
%
%
%
		
		%
		%
		
		%
		%
		%
		%
		%
		%
		%
		%
		%
		%
		%
		%
		%
		%
		%
		%
		%
		%
		%
		%
		%
		%
		%
		%
		%
		%
		%
		%
		%
		%
		%
		%
		%
		%
		%

		%
		%
		%
		%
		%
		%
		%

		
	\end{thebibliography}
	

\end{document}